\documentclass[letter]{aa} 

\usepackage{graphicx}
\usepackage{natbib}
\bibpunct{(}{)}{;}{a}{}{,} 
\usepackage[varg]{txfonts}
\usepackage{stfloats}
\usepackage{xurl}
\usepackage[colorlinks=true,linkcolor=bluea,allcolors=blue]{hyperref}

\def\o3{[O\textsc{iii}]}
\def\h2{H$_{\rm 2}$}

\def\msol{\,M$_{\rm \odot}$}
\def\mout{\,M$\rm _{\odot}\, yr^{-1}$}
\def\cm3{$\rm cm^{-3}$}

\def\s3{[S\textsc{iii}]}

\newcommand{\kms}{\,\hbox{\hbox{km}\,\hbox{s}$^{-1}$}}
\newcommand{\casa}{\textsc{CASA}}
\newcommand{\barolo}{\textsuperscript{3D}BAROLO}
\newcommand{\pajet}{PA$\rm_{jet}$}
\newcommand{\padisc}{PA$\rm_{disc}$}
\newcommand{\degree}{\ensuremath{^\circ}}
\newcommand{\ergs}{\,\hbox{\hbox{erg}\,\hbox{s}$^{-1}$}}
\newcommand{\ratio}{\hbox{T${\rm_{32}}$/T${\rm_{21}}$}}

\begin{document} 

   \title{Jet-induced molecular gas excitation and turbulence in the Teacup}
  \author{A. Audibert \inst{1,2}  \and C. Ramos Almeida \inst{1,2} \and S. Garc{\'i}a-Burillo\inst{3} \and F. Combes\inst{4} \and M. Bischetti\inst{5} \and M. Meenakshi\inst{6} \and D. Mukherjee\inst{6} 
 \and G. Bicknell\inst{7} \and A. Y. Wagner\inst{8} }

\institute{Instituto de Astrof\' isica de Canarias, Calle V\' ia L\'actea, s/n, E-38205, La Laguna, Tenerife, Spain  \email{anelise.audibert@iac.es} \and
 Departamento de Astrof\' isica, Universidad de La Laguna, E-38206, La Laguna, Tenerife, Spain \and
Observatorio Astron{\'o}mico Nacional (OAN-IGN)-Observatorio de Madrid, Alfonso XII, 3, 28014 Madrid, Spain \and
Observatoire de Paris, LERMA, Coll{\`e}ge de France, PSL University, CNRS, Sorbonne University, Paris, France \and 
Dipartimento di Fisica, Sezione di Astronomia, Universit\'a di Trieste, via Tiepolo 11, 34143 Trieste, Italy \and 
Inter-University Centre for Astronomy and Astrophysics, 
Savitribai Phule Pune University Campus, Pune- 411007, India \and 
Research School of Astronomy and Astrophysics, The Australian National University, Canberra, ACT 2611, Australia \and
Center for Computational Sciences, University of Tsukuba, Tennodai 1-1-1, 305-0006, Tsukuba, Ibaraki, Japan}

   \date{Received January 20, 2023; accepted February 26, 2023}

 
\abstract{In order to investigate the impact of radio jets on the interstellar medium (ISM) of galaxies hosting active galactic nuclei (AGN), we present subarcsecond resolution Atacama Large Millimeter/submillimeter Array (ALMA) \hbox{CO(2-1)} and \hbox{CO(3-2)} observations of the Teacup galaxy. This is a nearby ($D_{\rm L}$=388 Mpc) radio-quiet type-2 quasar (QSO2) with a compact radio jet ($P_{\rm jet}\approx$10$^{43}$\ergs) that subtends a small angle from the molecular gas disc. 
Enhanced emission line widths perpendicular to the jet orientation have been reported for several nearby AGN for the ionised gas. For the molecular gas in the Teacup, not only do we find this enhancement in the velocity dispersion but also a higher brightness temperature ratio (\ratio) perpendicular to the radio jet compared to the ratios found in the galaxy disc. Our results and the comparison with simulations suggest that the radio jet is compressing and accelerating the molecular gas, and driving a lateral outflow that shows enhanced velocity dispersion and higher gas excitation. These results provide further evidence that the coupling between the jet and the ISM is relevant to AGN feedback even in the case of radio-quiet galaxies.
}


   \maketitle
%

\section{Introduction}

The idea that outflows are almost ubiquitous in galaxies hosting active galactic nuclei (AGN) is supported by observational efforts in the past decades, as well as by the development of sub-grid physics in cosmological models \citep{cielo18, nelson19}. The inclusion of AGN feedback in the recipes of simulations is necessary to explain the observed bright end of the galaxy luminosity function, as otherwise galaxies would grow too big and massive \citep{croton06,dubois16}.

One of the potential drivers of multi-phase outflows, even in the case of radio-quiet AGN, are jets launched by the supermassive black hole (SMBH). These jets can strongly impact \hbox{(sub-)kpc} scales, by altering the properties of the interstellar medium (ISM) of the host galaxies. Observational evidence for outflows driven by jets has been reported in ionised \citep[e.g.,][]{heckman84,jarvis19,cazzoli22,girdhar22,speranza22} and molecular gas \citep[e.g.,][]{morganti15,oosterloo19,santi19,murthy22}. Supporting the observational evidence, hydrodynamic simulations \citep{wagner11,wagner12, mukherjee18sim} are able to reproduce the jet-ISM interaction and the impact of feedback induced by relativistic AGN jets inside the central kpc of gas-rich radio galaxies. The models trace the evolution of a relativistic jet propagating in a two-phase ISM, gradually dispersing the clouds through the effect of the ram pressure and internal energy of the non-thermal plasma, and creating a cocoon of shocked material that drives multi-phase outflows as the bubble expands.

In recent work, \citet[][hereafter {\color{blue}RA22}]{cra22} reported Atacama Large Millimeter/submillimeter Array (ALMA) \hbox{CO(2-1)} observations at $\sim$0\farcs2 (400\,pc) resolution of 7 radio-quiet type-2 quasars (QSO2s, i.e., obscured quasars) at redshifts z$\sim$0.1, with bolometric luminosities of L$_{\rm bol}\approx$10$^{\rm 45-46}$\ergs. These QSO2s are part of the Quasar Feedback (\href{http://research.iac.es/galeria/cra/qsofeed/}{QSOFEED}) sample. Cold molecular outflows with intermediate properties between those of Seyfert galaxies and ultra-luminous infrared galaxies (ULIRGs) were detected in the five QSO2s with CO(2-1) detections. Their molecular mass outflow rates are lower than those expected from their AGN luminosities \citep{fiore17,fluetsch19}, suggesting that other factors such as the jet power, the spatial distribution of the dense gas, and the coupling between jets and the dense gas, might also be relevant.

\begin{figure*}
\centering
 \includegraphics[width=17cm]{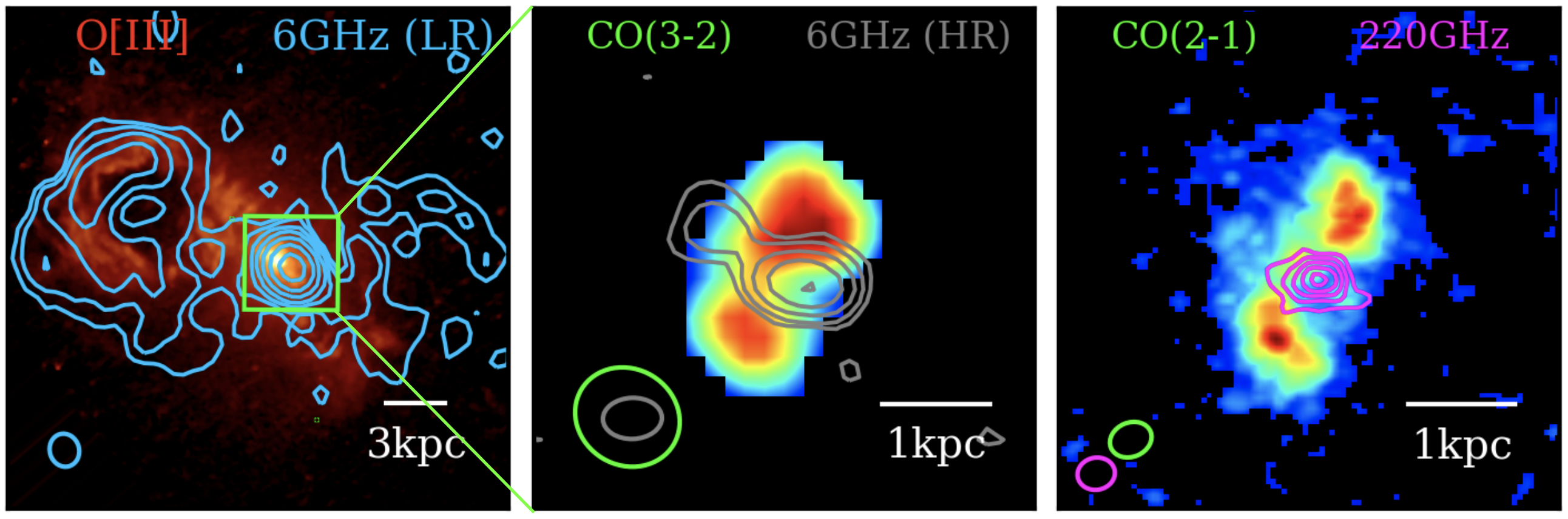}
\caption{The Teacup as seen in \o3, radio continuum, and CO. The left panel shows the large scale (15\arcsec$\times$15\arcsec) [O\textsc{iii}]$\lambda$5007\,\AA~HST image in colour with the low-resolution (LR) VLA 6\,GHz continuum contours overlaid in blue. The middle panel shows the 2\farcs8$\times$2\farcs8 peak intensity map of \hbox{CO(3-2)} with the high-resolution (HR) VLA 6\,GHz continuum contours in gray. The right panel corresponds to the \hbox{CO(2-1)} peak intensity map with the ALMA 220\,GHz continuum contours overlaid in magenta. The beam sizes are shown in the bottom left corner of each panel with corresponding colours.}
\label{fig:hst}
\end{figure*}

Among the QSO2s studied in {\color {blue} RA22}, the Teacup (SDSS\,J143029.88+133912.0; J1430+1339) revealed a peculiar CO morphology and disturbed kinematics. This, in addition to the compact radio jet (extent of $\sim$0.8\,kpc, PA=60\degree) detected in Very Large Array (VLA) data at 0\farcs25 resolution \citep{harrison15}, makes this object a promising target to study the jet-ISM interaction. It is hosted in a bulge-dominated galaxy showing signatures of a recent merger, including stellar shell-like features seen in the \textit{Hubble Space Telescope} (HST) images \citep{keel15}. The nickname of the Teacup stems from the 12\,kpc ionized gas bubble \citep{keel12} that is also detected in the radio with the VLA \citep{harrison15}, both shown in the left panel of Figure~\ref{fig:hst}. The \o3 kinematics reveal a $\sim$1\,kpc scale nuclear outflow with maximum velocities of $\sim$-750 km~s$^{\rm -1}$  \citep{vm14,harrison14}. In the near infrared (NIR), blueshifted broad components with maximum velocities of -1100\kms~were measured  from the Pa$\alpha$ and [Si\,VI] emission lines, also extending $\sim$1\,kpc and oriented at PA$\sim$70\degree~\citep{cra17}. The projected orientation and extension of the nuclear ionised outflow almost coincide with the compact jet, suggesting that the latter could be driving the outflow \citep{harrison15}. Therefore, the Teacup is one of the few nearby luminous AGN whose outflows have been characterized in different gas phases (ionized, warm, and cold molecular). This, together with its compact radio jet, makes it an ideal target to advance in our understanding of how compact jets impact the multi-phase ISM of radio-quiet AGN.

Here we present new ALMA \hbox{CO(3-2)} observations at $\sim$0\farcs6 (960\,pc) resolution, in addition to the \hbox{CO(2-1)} 0\farcs2 data from {\color{blue} RA22}, to investigate the influence of the jet on the molecular gas excitation and kinematics of the Teacup. We adopt a flat $\Lambda$CDM cosmology with H$_{\rm 0}$=70\kms Mpc$^{-1}$, $\Omega_{\rm M}$=0.3, and $\Omega_{\rm \Lambda}$=0.7.

\section{Observations and data reduction}\label{sec:data}

We present ALMA observations of the \hbox{CO(2-1)} and \hbox{CO(3-2)} emission lines observed in bands 6 and 7. The \hbox{CO(2-1)} observations and data reduction are described in detail in {\color {blue} RA22}. The \hbox{CO(2-1)} datacube has a synthesised beam size of 0\farcs21$\times$0\farcs18 and a root mean square (rms) noise of 0.39\,mJy/beam per channel of 10\kms. The \hbox{CO(3-2)} data (2016.1.01535.S; PI: G. Lansbury) were retrieved from the ALMA archive. The observations were done in April 2017, in the C40-3 configuration, with an on-source integration time of 30.3 min and in a single pointing, covering a field-of-view (FoV) of 18\arcsec. The spectral window of 1.875\,GHz total bandwidth was centred at the \hbox{CO(3-2)} line, with a channel spacing of 7.8\,MHz, corresponding to $\sim$7.2\kms~after Hanning smoothing. 

The \hbox{CO(3-2)} data were calibrated using the \casa~ software version 4.7.2 \citep{casa} in the pipeline mode. As flux, bandpass, and phase calibrator the sources J1550+0527, J1337-1257 and J1446+1721 were used. Once calibrated, the imaging and cleaning were performed with the task \textsc{tclean}. The spectral line map was obtained after subtracting the continuum in the $uv$-plane using the tasks \textsc{uvcontsub} and a 0th order polynomial in the channels free from emission lines. The \hbox{CO(3-2)} data cube was produced with a spectral resolution of 10\kms~(10.6\,MHz) and using Briggs weighting mode and a robust parameter set to 0.5 in order to achieve the best compromise between resolution and sensitivity. Finally, the datacube was corrected for primary beam attenuation, resulting in a synthesised beam of 0\farcs60 $\times$0\farcs54 at PA=59\degree~and a rms noise of 1.9\,mJy~beam$^{-1}$ per 10\kms~ channel. In order to study the line ratios and compare the \hbox{CO(2-1)} to the new \hbox{CO(3-2)} datacube, we regridded the \hbox{CO(2-1)} data to the same pixel scale of the \hbox{CO(3-2)} and then convolved it to the same common beam size of 0\farcs60$\times$0\farcs54 (960\,pc$\times$860\,pc).

The auxiliary radio data used here are the 6\,GHz VLA observations of the Teacup in C-band in two different configurations: the high-resolution (0\farcs25, HR) A-array and the low-resolution (1\farcs0, LR) B-array maps. The data were presented and analysed in full by \citet{jarvis19}.

\section{Results and analysis}\label{sec:results}

The \hbox{CO(3-2)} peak intensity map, shown in the middle panel of Figure~\ref{fig:hst}, 
shows the double peaked morphology first detected in \hbox{CO(2-1)} and discussed in {\color {blue} RA22}.  
The distance between the two peaks is $\sim$0\farcs8 (1.3\,kpc). 
The integrated intensity, intensity weighted velocity field, and velocity dispersion ($\sigma$) maps of the \hbox{CO(3-2)} emission are shown in Appendix \ref{appendixA}. Overall, the 1st and 2nd moment maps (velocity field and $\sigma$) are similar to their CO(2-1) counterparts, although with less level of detail due to the coarser angular resolution. For example, the velocity field shows the disc rotation pattern, but the disturbed kinematics seen in \hbox{CO(2-1)} are not detected in \hbox{CO(3-2)}. 
The 2nd order moment map shows high values of $\sigma$, reaching $\sim$120\kms~with PA$\sim$-30\degree, orthogonal to the radio jet orientation (PA$\rm_{jet}$=60\degree), similar to the $\sigma$ map of the \hbox{CO(2-1)} data in {\color {blue} RA22}. 

\begin{figure*}
\centering
\includegraphics[width=17cm]{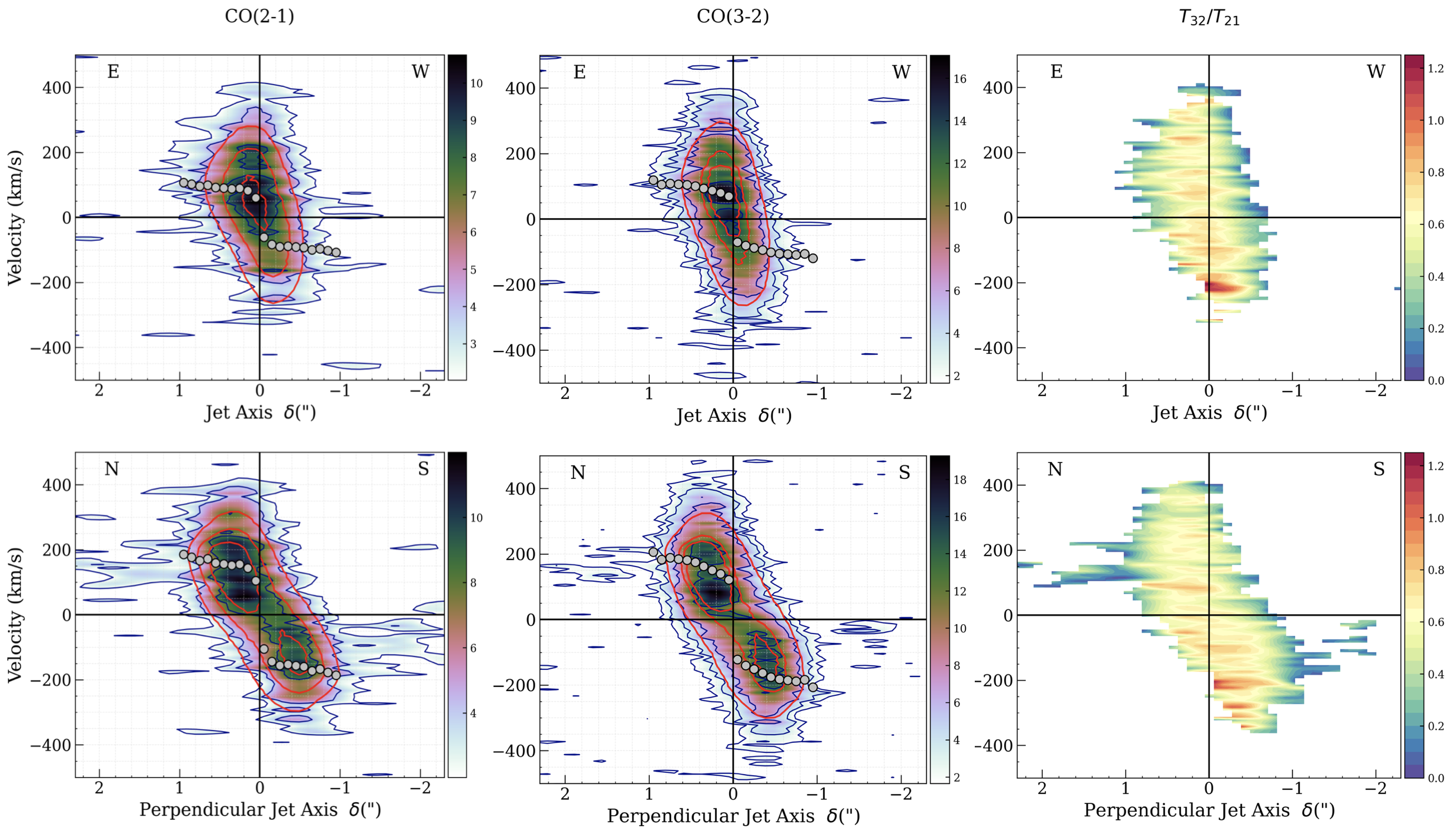}
\caption{PVDs along (PA=60\degree) and perpendicular (PA=-30\degree) to the radio jet direction, extracted in a slit of 0\farcs6 width and showing CO emission. 
The contours are 2, 4, 8, and 16$\sigma_{\rm rms}$, with $\sigma_{\rm rms}$=1\,mJy~beam$^{-1}$ (left panel) for CO(2-1) and $\sigma_{\rm rms}$=1.9\,mJy~beam$^{-1}$ (middle panel) for CO(3-2). The red contours shown in the left and middle panels correspond to the corresponding \barolo~ model at 2, 4 and 6$\sigma_{\rm rms}$, and the grey dots are the derived rotation velocities projected to the respective PA. 
Right panels are the PVDs of the \hbox{CO(3-2)/CO(2-1)} line ratios. The colour bars indicate the scales (flux density in mJy for the \hbox{CO(3-2)} and \hbox{CO(2-1)} PVDs and dimensionless in brightness temperature ratio \ratio).}
\label{fig:pvd}
\end{figure*}

\begin{figure*}
\centering
\includegraphics[width=17cm]{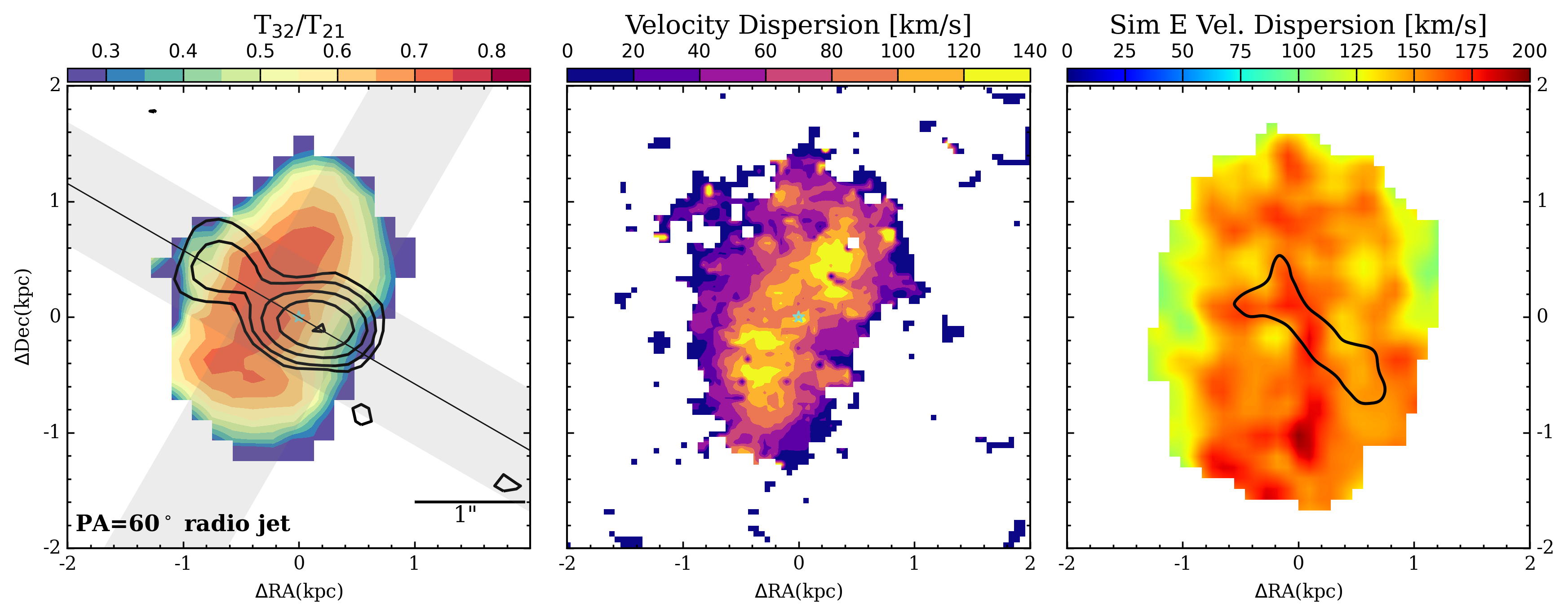}
\caption{Brightness temperature ratio (\ratio) and CO(2-1) velocity dispersion maps. The left panel shows the \ratio~map at 0\farcs6 resolution with the VLA 6\,GHz HR contours overlaid in black (at 4, 8, 16, 32, and 64$\sigma_{\rm rms}$, with $\sigma_{\rm rms}$=23\,$\mu$Jy/beam). The \pajet=60\degree~is indicated as a black solid line. The orientation and width (0\farcs6) of the slits used to produce the PVDs shown in Figure~\ref{fig:pvd} are indicated as grey shaded areas. The middle panel corresponds to the $\sigma$ map of the CO(2-1) line emission at 0\farcs2 resolution. The cyan star in the left and middle panels indicates the peak of the 220\,GHz ALMA continuum. On the right panel, we show the $\sigma$ map of the gas from Simulation E in \citet{meenakshi22} for a jet power of $P_{\rm jet}=10^{45}$\ergs, inclined 20\degree~from the CO disc plane, and with an angle of $\sim$25\degree~relative to the LOS.}

\label{fig:t32}
\end{figure*}

In order to investigate the possible influence of the jet on kinematics of the molecular gas, 
we created position-velocity diagrams (PVDs) along and perpendicular to the direction of the jet axis. We extracted them using a slit width of 0\farcs6, equivalent the common resolution of the \hbox{CO(3-2)} and convolved \hbox{CO(2-1)} cubes. The PVDs are shown in Figure~\ref{fig:pvd} and the position and orientation of the slits are indicated as shaded rectangles in the left panel of Figure~\ref{fig:t32}.  A broad gradient of velocities is detected both along and perpendicular to the jet axis, spanning up to $\pm$400\kms~for both the \hbox{CO(2-1)} and \hbox{CO(3-2)} emission.
We detect gas at higher velocities in the PVDs of \hbox{CO(3-2)} than in those of \hbox{CO(2-1)}, specially in the case of the blueshifted velocities.
The PVDs perpendicular to the radio jet show an inverted S-shape in the central $\sim$2\arcsec~(3.2 kpc), which includes the bulk of the emission. 
There is also a contribution from slow velocities ($<$200\kms) at $r>$1\arcsec~that
resemble a rotation pattern, unlike the central part of the PVDs. These features were not reported in {\color{blue} RA22} because of the narrower slit width and different orientations there considered.

We analysed the \hbox{CO(2-1)} and \hbox{CO(3-2)} kinematics at 0\farcs6 resolution using the ``3D-Based Analysis of Rotating Objects from Line Observations'' (\barolo) software by \citet{barolo15}, following the methodology described in {\color {blue} RA22}. For the 3DFIT, we applied a mask using a threshold of 3$\sigma_{\rm rms}$. We fixed the PA and the inclination of the disc to the values derived using the high-resolution \hbox{CO(2-1)} data in {\color {blue} RA22}: \padisc=4\degree and $i$=38\degree, and the centre position was fixed to the continuum peak at 220GHz (RA=14h30m29.88s and Dec=+13d39m11.93s). Then, we allowed \barolo~to fit the rotation velocity 
and velocity dispersion. From the model datacube created by \barolo, we produced PVDs extracted in slits with the same width and orientations described above, and
overlaid them to the \hbox{CO(2-1)} and \hbox{CO(3-2)} PVDs (see red contours in Figure \ref{fig:pvd}). The bulk of the \hbox{CO(2-1)} and \hbox{CO(3-2)} emission at 0\farcs6 resolution can be well reproduced by our \barolo~model. We find 77\% and 90\% of gas in rotation in \hbox{CO(2-1)} and \hbox{CO(3-2)}, respectively, with a central velocity dispersion of $\sim$100\kms. For comparison, in the case of the \hbox{CO(2-1)} data at 0\farcs2 resolution, only 55\% of the gas can be accounted for by rotation. This is because a significant fraction of  non-circular and tangential motions are diluted due to the larger beam size of the CO(3-2) data.  
However, as shown in Fig. \ref{fig:pvd}, the molecular gas with high velocities ($\sim\pm$400\kms) cannot be explained by rotation, as the model reproduces the rotation pattern up to maximum values of around $\pm$300\kms.

The right panels of Figure~\ref{fig:pvd} show the PVDs of the \hbox{CO(3-2)/CO(2-1)} line ratio. 
To produce these PVDs, the units of the CO(3-2) and CO(2-1) PVDs were first converted to brightness temperature (T$\rm_{32}$ and T$\rm_{21}$, in units of K), 
considering only CO emission above 2$\sigma_{\rm rms}$. A clear increase in the line ratio is seen in regions of 0\farcs4 radius ($\sim$600\,pc) along and perpendicular to the radio jet, up to values of 1.2. 
In order to study this line ratio in a spatially-resolved manner, we created the \ratio~line ratio of the integrated intensity (moment 0) maps, which is shown in the left panel of Figure~\ref{fig:t32}. This line ratio map shows the highest values, of $\sim$0.8, along the direction perpendicular to the radio jet, while
other regions of the disc present ratios of 0.3$<T\rm_{32}/T\rm_{21}<$0.5, similar to typical values found in settled discs or in the Milky Way \citep{leroy09, carilli13}. We note that the values of \ratio~in the integrated line ratio map shown in Figure~\ref{fig:t32} are lower than those in the PVDs shown in Figure~\ref{fig:pvd} because the former are average values along the LOS.
In the right panel of Figure~\ref{fig:t32} we also show the $\sigma$ map of the original \hbox{CO(2-1)} data (i.e., at 0\farcs2 resolution). The region with larger $\sigma$ values (i.e., highest turbulence) coincides with that having the highest values of \ratio. As discussed in Section \ref{discussion}, enhanced \ratio~has been reported for a few nearby radio-loud AGN along the direction of the jet (e.g., \citealt{dasyra16,oosterloo17,oosterloo19}), but to the best of our knowledge, this is the first detection of increased \ratio~along the direction {\it perpendicular} to the radio jet in a radio-quiet AGN.

Using the 1.5\,GHz VLA flux and the spectral index of $\alpha$=-0.87 reported by \citet{jarvis19}, it is possible to estimate the jet power of the Teacup from the relations of e.g., \cite{birzan08} and \cite{cavagnolo10}. If we do so, we obtain $P_{\rm jet}\simeq$1-3$\times$10$^{43}$\ergs. However, these scaling relations have large scatter ($>$0.78\,dex) and are drawn from samples of galaxies dominated by more evolved jets. Therefore, they may not apply to compact jets confined into the galaxy ISM as it is the case of the Teacup. As pointed out by \cite{godfrey16}, the $P_{\rm jet}$-${L_{\rm radio}}$ relations may be quite unreliable because, among other things, they neglect the effect of distance, morphology, and environment. Here we do not attempt to derive an accurate measurement of the jet power, we are only interested in its order of magnitude, as we aim at comparing with the simulations described in Section \ref{sec:sim}.

To perform this comparison, we also need a rough approximation of the jet inclination relative to the CO disc. Although it is not possible to infer a precise angle using the radio data available, we obtain an estimate by using the core to extended radio flux ratio parameter, $R_{\rm c}=\frac{S_{\rm core}}{S_{\rm ext}}$. $S_{\rm core}$ is the core flux density of the unresolved beamed nuclear jet, and $S_{\rm ext}$ is the flux density of the unbeamed lobes. Statistically, $R_{\rm c}$ should be larger in more beamed objects, but the extended flux is time-dependent and it depends on the frequency of observation. In the case of the Teacup, using the flux densities at 5.2\,GHz reported by \citet{jarvis19} for the HR VLA data, $R_{\rm c}$=(2.3/0.6)=3.8, which would correspond to an angle of $\sim$20\degree~relative to the line-of-sight (LOS) if we use Eq. 1 in \citet{orr82} with $\gamma$=5 and $R_{\rm T}$=0.024. Considering the inclination of the CO disc derived from the modelling with \barolo ~ ($i_{\rm CO}$=38\degree; i.e., 52\degree~relative to the LOS; {\color {blue} RA22}), we conclude that the jet likely subtends a small angle relative to the CO disc. The results presented here suggest that the radio jet is strongly coupled to the CO disc and because of that, it is driving turbulence and exciting the molecular gas along PA$\sim$-30\degree.

\section{Discussion}\label{discussion}

In this Letter we present robust observational evidence that the radio jet is interacting with the cold ISM of the Teacup and causing the disturbed kinematics, large turbulence, and peculiar excitation conditions of the gas. In the following, we discuss the results and compare them with high resolution simulations of jets propagating on the ISM of galaxies.

\subsection{Physical conditions and kinematics of the molecular gas}

Spatially resolved CO emission line ratio analyses have been reported for a few nearby Seyferts and radio galaxies. \citet{viti14} modelled the excitation conditions of the molecular gas in the circumnuclear disc of NGC\,1068, and claimed that elevated values of \hbox{$R_{\rm 31} \equiv T_{\rm32}/T_{\rm 10}\sim$5} (i.e., involving the \hbox{CO(3-2)} and \hbox{CO(1-0)} emission lines) and \hbox{$R_{\rm 21} \equiv T_{\rm21}/T_{\rm 10}\gtrsim$4} (involving \hbox{CO(2-1)} and \hbox{CO(1-0)} lines) correspond to hot dense gas (kinetic temperatures $T>$150\,K and $\eta>10^5$\,cm$^{-3}$) excited by the AGN. 
Higher values of different emission line ratios have been reported for the molecular gas interacting with the jet in the radio galaxy PKS\,1549-79 \citep{oosterloo19}, and also in the Seyfert galaxy IC\,5063 \citep{dasyra16,oosterloo17}. This indicates that jet-impacted regions have different gas excitation conditions due to shocks and/or reduced optical thickness \citep{morganti21}. IC\,5063 is a clear-cut example of jet coplanar with the galaxy disc, which is driving a multi-phase outflow \citep{tadhunter14,morganti15}. The jet power estimated for IC\,5063 is $P_{\rm jet}\sim$10$^{\rm 44}$\ergs~and the PVD along the jet axis revealed \ratio~values ranging from 1.0$-$1.5 in the outflowing regions, clearly different from the gas following regular rotation ($R_{\rm 32}<0.7$). Under local thermodynamic equilibrium (LTE) conditions, $R_{\rm 32}>$1 corresponds to excitation temperatures ($T_{\rm ex}$) of $\sim$50\,K \citep{oosterloo17}. A similar temperature was found for the fast outflow gas component of IC\,5063 using \hbox{$R_{\rm 42}\equiv T{\rm_{43}}/T{\rm_{21}}>$1} (i.e., involving the CO(4-3) and CO(2-1) lines), suggesting that the gas in the outflow is optically thin \citep{dasyra16}. 

The Teacup presents similar values of \ratio~as those reported in the outflow region of IC\,5063 (i.e., \ratio$>$1; see Figure~\ref{fig:pvd}). The difference is that in the Teacup, the highest values are found perpendicular and not along the jet direction, as shown in Fig.~\ref{fig:t32}. This is indicative of the presence of hot dense gas, with $T_{\rm ex}\sim$50\,K, probably excited by the cocoon of shocked gas that is driving the lateral outflow/turbulence that we observe in CO.

Regarding the gas kinematics, examples of enhanced $\sigma$ of the ionised gas along the direction perpendicular to jet axis have already been reported in the literature for nearby AGN. One example is the radio galaxy 3C\,33, for which \citet{couto17} proposed a scenario in which the ISM is laterally expanding due to the passage of the jet, supported by the optical emission line ratios that they measured in the same region, typical of gas excitation induced by shocks. 
\citet{venturi21} reported high [O\textsc{iii}] velocity widths in four nearby Seyfert galaxies from the MAGNUM survey hosting low power radio jets. These high $\sigma$ values were interpreted as due to the passage of a low inclination jet through the galaxy discs ISM, which produces shocks and induces turbulence.
The MURALES survey of nearby 3CR radio sources has revealed similar features in [O\textsc{iii}] for some of the radio galaxies \citep{balmaverde19,balmaverde22}. 
In the Teacup, we measure the highest values of the \hbox{CO(2-1)} and \hbox{CO(3-2)} $\sigma$, of up to 140\kms, perpendicular to the jet direction. We note that the enhancement in \ratio~is shifted by $\sim$0\farcs2 to the east relative to the gas with the highest $\sigma$ (see Figures \ref{fig:t32} and \ref{fig:mom}). High angular resolution CO(3-2) observations are required for exploring this shift in more detail.

\subsection{Is the radio jet driving a lateral molecular outflow?}
\label{sec:sim}

Simulations show that strong jet-induced feedback can occur even when $P_{\rm jet}\sim$10$^{\rm 43-44}$\ergs~\citep{mukherjee16,talbot22}. These low-power jets are trapped in the galaxy’s ISM for longer times and therefore they are able to disrupt the ISM over a larger volume than more powerful jets \citep{nyland18}. 
Another crucial element for having efficient feedback is the jet-ISM coupling. Simulations show that stronger coupling occurs when jets have low inclination relative to the gas disc, since jets are confined within the galaxy ISM, injecting more kinetic energy into the surrounding gas and inducing local turbulence and shocks \citep{mukherjee18sim, meenakshi22}. According to the estimate described in Section \ref{sec:results}, this seems to be the case of the Teacup.

We can compare the PVDs shown in Figure \ref{fig:pvd} with simulation E in \citet{meenakshi22}. This simulation consists of a jet of $P_{\rm jet}\sim$10$^{\rm 45}$\ergs~propagating in a gas disc with a central gas density of 200 cm$^{\rm -3}$, and subtending an angle of 20\degree~(i.e., almost coplanar). This jet power is higher than our estimate for the Teacup ($\sim$10$^{\rm 43}$\ergs), but in addition to the caveats mentioned in Section \ref{sec:results}, $P_{\rm jet}$ values estimated from the \citet{birzan08} and \citet{cavagnolo10} relations are usually lower limits for kpc-scale confined jets as it is the case of the Teacup. For instance, in the case of IC\,5063, the jet power required to model the jet-induced molecular gas dispersion was a factor of 3 to 10 higher than that inferred from scaling relations \citep{mukherjee18ic}.
The jet does not only induce turbulence due to the injection of energy within its immediate vicinity but also causes a large broadening in the PVDs, with velocities reaching up $\sim$400\kms~in the regions both along and perpendicular to the jet path (see Figs. 8 and 10 in \citealt{meenakshi22}).

Motivated by the resemblance between the ionised gas features in Sim E and the molecular gas features in the Teacup, we performed a comparison between our observations and the simulations in \citet{meenakshi22}. Although the impact of jets on the dense molecular gas is not directly considered in the simulations, the emissivity of the CO(2-1) gas can be approximated with functions that depend on gas density, temperature, and cloud tracers, all extracted from the simulations. This method is similar to that adopted in \citet{mukherjee18ic} to reproduce the molecular gas features observed in IC\,5063. The details of the simulations are presented in Appendix~\ref{simulations}. As shown in the right panel of Figure~\ref{fig:t32}, the morphology of the gas velocity dispersion predicted by Sim E is similar to the Teacup, showing high values close to perpendicular to the jet direction. This happens because the main jet stream, as it propagates through the clumpy ISM, is split and deflected multiple times, carrying momentum in all directions including directions perpendicular to the jet axis and accelerating gas out of the disc plane. The simulated \hbox{CO(2-1)} PVDs along and perpendicular to the jet (see Figure~\ref{fig:pv_sim}) show high velocity components reaching $\pm$400\kms~within the central 2 kpc of the galaxy. Even though these simulations were not tailored to model the potential and radio power of the Teacup, they reproduce qualitatively the main kinematic features of the \hbox{CO(3-2)} and \hbox{CO(2-1)} observations.

\subsection{Molecular outflow scenarios}

Based on the CO(2-1) data at 0\farcs2 resolution, {\color{blue} RA22} reported evidence for a molecular outflow, coplanar with the CO disc, with a velocity of 185\kms, radius of 0.5\,kpc, and deprojected outflow rate of $\dot{M}_{\rm out}$=15.8\,M$_{\odot}$~yr$^{\rm -1}$. This outflow rate is most likely a lower limit in the case of the Teacup because of the complex CO kinematics, and because {\color{blue} RA22} only considered non-circular gas motions along the CO disc minor axis to compute the outflow mass. 
 However, the PVDs along and perpendicular to the jet axis in Figure~\ref{fig:pvd}  
show high velocities, of up to $\pm$400\kms, that cannot be reproduced with rotation, and both \ratio~and $\sigma$ (see Figure~\ref{fig:t32}) show higher values in the direction perpendicular to the jet (i.e., closer to the CO major axis). Therefore, it seems plausible that the higher gas excitation and increased turbulence that we find along PA=-30\degree~are signatures of a jet-driven molecular outflow. We then considered different scenarios to compute the molecular outflow mass, which varies between 3$\times$10$^7$ and 6$\times$10$^8$ M$_{\sun}$. The details are given in Appendix \ref{appendixC}.

Scenario \textsc{i}, the least conservative, consists of assuming that all the molecular gas that is not rotating, at least according to our \barolo~model, is outflowing. Based on this model, we find that half of the molecular gas mass is rotating, and the mass outflow rate is 44\mout~using an average  outflow velocity of 100\kms. 
Scenario \textsc{ii} assumes that only the high-velocity gas (faster than $\pm$300\kms) participates in the outflow, and for this we measure 41\mout. Finally, in Scenarios \textsc{iii} and \textsc{iv} we measure outflow rates of 15.1 and 6.7\mout~by subtracting the rotation curve from the \hbox{CO(2-1)} datacube and then considering high-velocity gas only (faster than $\pm$300\kms~in Scenario \textsc{iii} and just the high-velocity wings shown in Figure \ref{fig:res} in Scenario \textsc{iv}). 

Based on the results presented in Section \ref{sec:results}, aided by the simulations discussed in Section \ref{sec:sim}, we favour a scenario where the jet is pushing the molecular gas out of the disc plane and producing a lateral expansion that compresses the gas, increasing turbulence and gas excitation. 
Therefore, in principle we favour Scenarios \textsc{ii} and \textsc{iii}, which consider high-velocity gas only, without specifying any particular direction for the outflow. Considering this, the outflow mass would be in the range 5.6--16$\times$10$^7$ M$_{\sun}$, and the outflow mass rate within 15-41\mout.

Jets produce efficient feedback by increasing the turbulence of the dense gas when the ratio between the radio power and the Eddington luminosity, P$_{\rm jet}$/L$_{\rm Edd}>$10$^{-4}$ \citep{wagner12}. In the Teacup, this ratio is $\sim$9$\times$10$^{-4}$, and the jet power is enough to drive the molecular outflow because it is about two or three orders of magnitude higher than the outflow kinetic power (see Table \ref{tab:prop}). However, the fate of the molecular gas during this feedback episode is a galactic fountain, since the escape velocity of the galaxy v$_{\rm esc}\gtrsim$530\,\kms, estimated from its dynamical mass as in \citet{feruglio20}, is larger than the fastest velocities in the outflow ($\sim\pm$400\,\kms). Regardless, this will have a global effect on the galaxy’s evolution by redistributing mass, metals, and delaying further star-formation.

Based on the results presented here, together with the predictions from the models described in Section \ref{sec:sim}, jet-induced turbulence and shocks are the most likely mechanisms to explain the high values of $\sigma$ and \ratio~reported here for the cold molecular gas in the Teacup. Larger samples of radio-quiet quasars showing low-power radio jets are required to better quantify the impact of jet-ISM coupling in the kinematics of the molecular gas.

\begin{acknowledgements}

We are very grateful to the referee for very helpful and detailed comments, which improved the presentation of the paper. AA and CRA acknowledge the projects ``Quantifying the impact of quasar feedback on galaxy evolution'', with reference EUR2020-112266, funded by MICINN-AEI/10.13039/501100011033 and the European Union NextGenerationEU/PRTR, and the Consejer\' ia de Econom\' ia, Conocimiento y Empleo del Gobierno de Canarias and the European Regional Development Fund (ERDF) under grant ``Quasar feedback and molecular gas reservoirs'', with reference ProID2020010105, ACCISI/FEDER, UE. AA acknowledges funding from the MICINN (Spain) through the Juan de la Cierva-Formación program under contract FJC2020-046224-I. SGB  acknowledges support from the research project PID2019-106027GA-C44 of the Spanish Ministerio de Ciencia e Innovación. This paper makes use of the following ALMA data: ADS/JAO.ALMA\#2018.1.00870.S, and ADS/JAO.ALMA\#2016.1.01535.S. ALMA is a partnership of ESO (representing its member states), NSF (USA) and NINS (Japan), together with NRC(Canada) and NSC and ASIAA (Taiwan), in cooperation with the Republic of Chile. The Joint ALMA Observatory is operated by ESO, AUI/NRAO and NAOJ. The National Radio Astronomy Observatory is a facility of the National Science Foundation operated under cooperative agreement by Associated Universities, Inc. 

\end{acknowledgements}

\bibliographystyle{aa} 
\bibliography{REF.bib}

\begin{appendix}

\section{Moment maps of the CO(3-2) emission}
\label{appendixA}

\begin{figure}[h]
 \resizebox{\hsize}{!}{\includegraphics{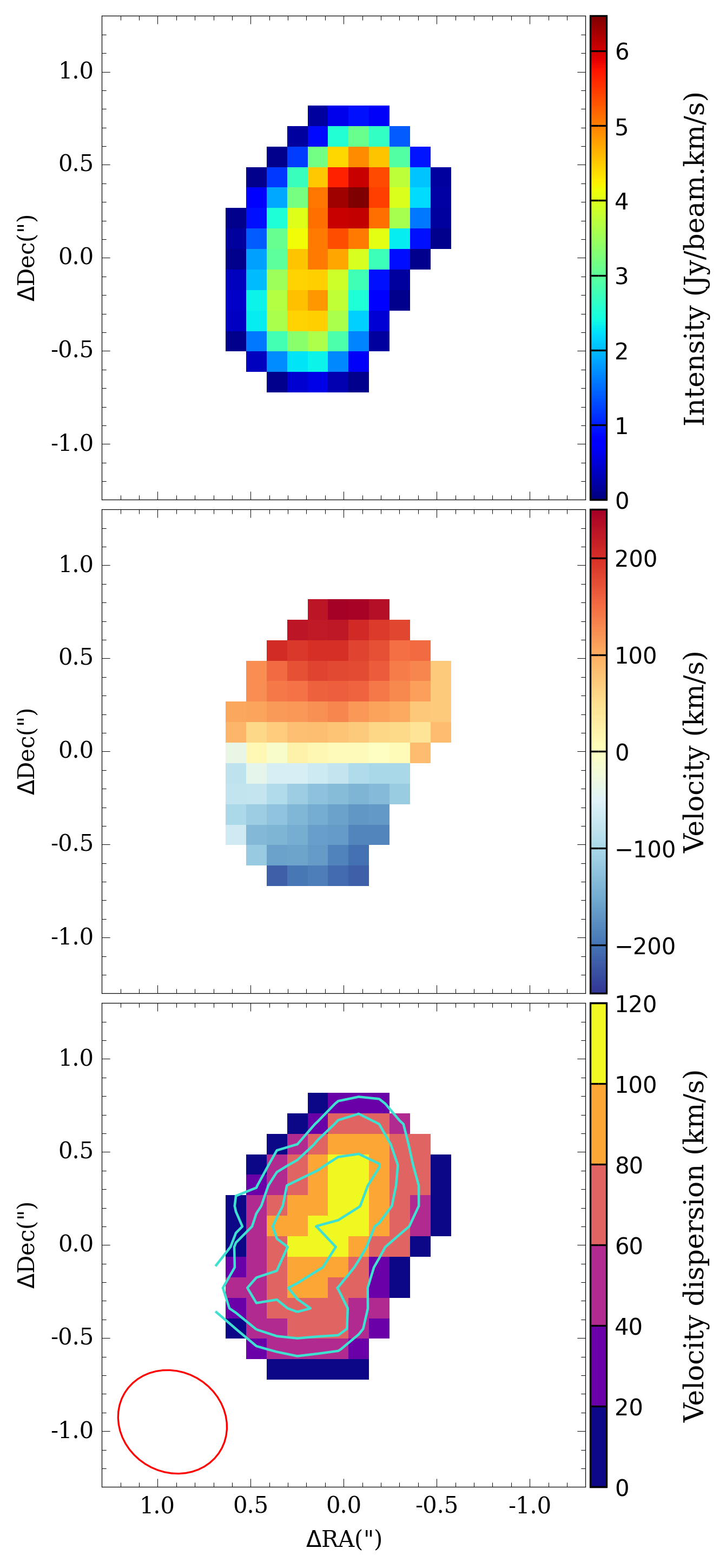}}
\caption{Moment maps of the CO(3-2) emission of the Teacup. Top, middle, and bottom panels correspond to the integrated intensity (moment 0, in Jy~beam$^{-1}$~km~s$^{-1}$ units), intensity weighted velocity field (moment 1, in \kms), and $\sigma$ (moment 2, in \kms), respectively. East is to the left and north to the top. In the $\sigma$ map, the contours show the \ratio~of Figure~\ref{fig:t32} at 0.5, 0.6, and 0.7, and the red ellipse in the bottom left corner indicates the beam size.}
\label{fig:mom}
\end{figure}

We computed the moment maps of the CO(3-2) emission at $\sim$0\farcs6 resolution as described in Section \ref{sec:results}. The integrated intensity, intensity-weighted velocity field, and velocity dispersion maps (moments 0, 1, and 2) are shown in Figure \ref{fig:mom}.

\section{Simulation of jet propagating in the molecular disc}\label{simulations}

\begin{figure*}
    \centering
    \includegraphics[width=17cm]{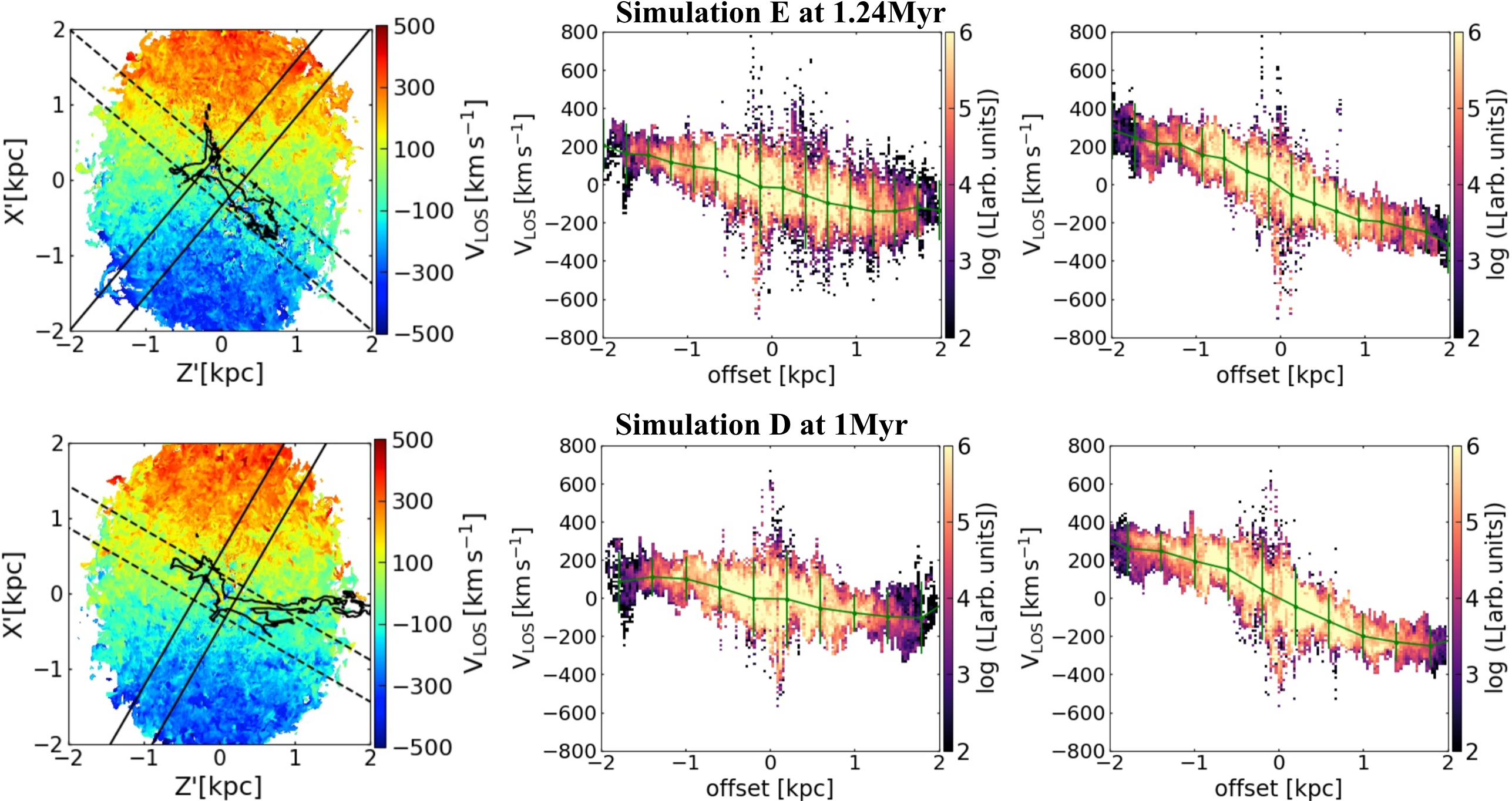}
    \caption{Snapshots of the mean velocity field and PVDs from Sim E (top panels, jet inclined 20\degree~relative to the disc) and Sim D (bottom panels, jet inclined at 45\degree) for $P_{\rm jet}$=10$^{45}$\ergs~at 1.24\,Myr and 1\,Myr, respectively. Left panels: mean LOS velocity fields with the position of the slits of 500\,pc width along (dashed lines) and perpendicular (solid lines) to the jet. Middle and right panels: PVDs along and perpendicular to the jet. The green curves are the mean velocity curve with $\pm$2$\sigma$ deviation, with a maximum deviations along and perpendicular to the jet reaching 408 (448)\kms~and 434 (412)\kms, respectively, for Sim E (Sim D).}
    \label{fig:pv_sim}
\end{figure*}

\begin{figure}[h]
 \resizebox{\hsize}{!}{\includegraphics{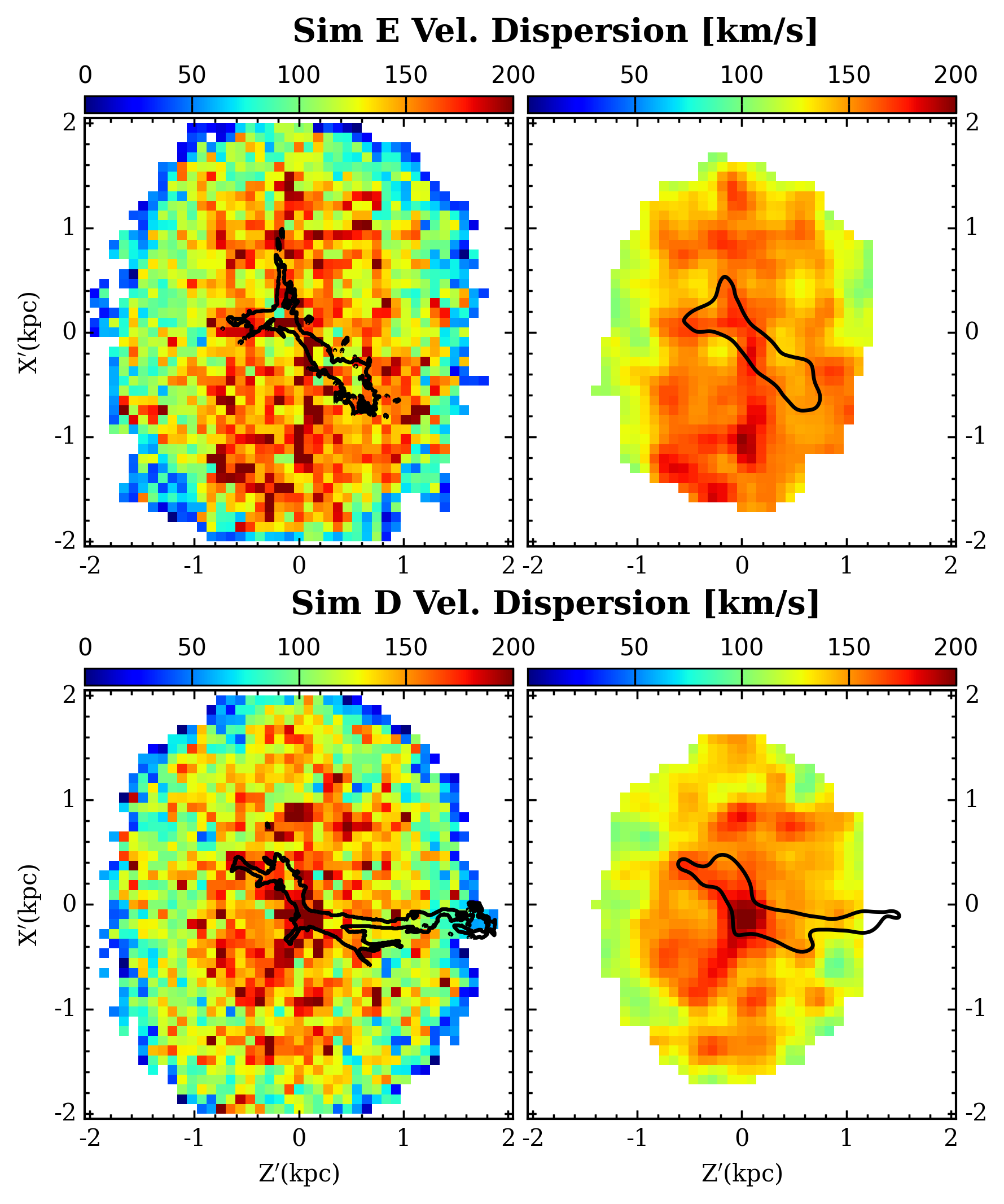}}
\caption{$\sigma$ maps from Sim E (top panels) and D (bottom panels). Left panels: maps at a resolution of 100\,pc. Right panels: simulated maps and jet-tracer smoothed by convolving with a 2D Gaussian of width 100 pc and 60 pc, respectively. The jet tracer is shown at 0.05 contour level.}
\label{fig:vdispSIMS}
\end{figure}

In order to elucidate the nature of jet-ISM interaction in the Teacup, we present here qualitative comparisons with the kinematics of the dense gas in simulations D and E of \citet[][hereafter Sim D and E]{mukherjee18sim} using the diagnostics developed in \citet{meenakshi22}. In these simulations, jets of power $10^{45}~\mbox{erg s}^{-1}$ are launched at angles of $45^\circ$ and $20^\circ$ from the disc plane for Sim~D and E respectively. These inclinations resemble the angle subtended by the jet and the CO disc in the Teacup. It should be noted that although the simulations are not tailored to mimic the Teacup,  we performed a qualitative comparison to try to understand the jet-gas interaction in this galaxy, as in \citet{murthy22}. 

In these simulations a turbulent dense gas disc of radius $\sim$2\,kpc is placed in the $X-Y$ plane. A pair of relativistic jets are launched from the centre. The jet's axis lie in the $X-Z$ plane, inclined from the disc at different angles of choice. According to the simulation's conventions, $\theta_{\rm I}$ = 90\degree, $\Phi_{\rm I}$ = 360\degree~corresponds to the edge-on view, with the observer facing the $X-Z$ plane. For the analysis below, we have used the value $\theta_{\rm I}=52\degree$ to match that of the Teacup. The azimuthal angle $\phi_{\rm I}$ is varied to match the observed relative orientation of the projected jet's axis to the disc. The image plane is inverted to the observed orientation of Teacup to facilitate the visual comparison of the simulated maps.

For the analysis, we constructed  mean LOS velocities, and the PVDs weighted by synthetic CO luminosities, following the method outlined in Section~5.2 of \citet{mukherjee18ic} and adopting the same parameters as in their work. Dense gas regions with temperatures lower than 5000\,K and densities
higher than 10\,$\mbox{cm}^{-3}$ were used for this analysis. The PVDs are then produced using slits of 500~pc in width oriented along and perpendicular to the jet. In Figure~\ref{fig:pv_sim}, the top left panel shows the projected velocity map for Sim E at a $\theta_{\rm I}=52\degree$ and $\phi_{\rm I}=340\degree$, overlaid by the jet tracer at a value of 0.05. At this orientation, the jet subtends an angle of 25\degree from the LOS. The PVDs along and perpendicular to the jet are shown in the top right panels. 
The jet induces high velocities along both slits, reaching values of up to $\sim\pm$400~\kms, and with the largest dispersion appearing in the central 1~kpc regions. The morphology of the PVD for the perpendicular slit is also very similar to that obtained for the Teacup in Figure~\ref{fig:pvd}, which appears to be sharply curved due to contribution from the rotation pattern of the disc. In the bottom panels of Fig.~\ref{fig:pv_sim}, we show the mean velocity field and PVDs for Sim~D, which are produced for an image plane oriented at $\theta_{\rm I}=52\degree$ and $\phi_{\rm I}=30\degree$. This corresponds to an angle of 23.5\degree~between the jet and the LOS. Similar to Sim~E, the PVDs along and perpendicular to the jet in Figure~\ref{fig:pv_sim} also exhibit velocities of up to $\sim\pm$400\kms~within the central 2~kpc.

The simulated $\sigma$ maps at the resolution of 100\,pc and smoothed in the 2D domain by convolving with a Gaussian of similar kernel width are shown in Figure~\ref{fig:vdispSIMS}.
The $\sigma$ distribution from Sim~E (top panels) displays the highest values along the direction perpendicular to the jet path, similar to the observed spatial distribution of $\sigma$, shown in Figure~\ref{fig:t32}. The $\sigma$ map from Sim~D (see bottom panels of Fig.~\ref{fig:vdispSIMS}) also shows high values in the regions perpendicular to jet, although more compact than in Sim~E, with the highest values along the jet. This extended velocity dispersion is caused by the jet-driven outflows along the minor axis, as well as along other directions throughout the disc, with the former being more prominent \citep[e.g., see Figs. 7 and 8 in][]{meenakshi22}. This indicates that the jet-gas interaction in the nuclear regions of the disc can lead to high-velocity dispersion in the molecular gas distribution that can extend in directions perpendicular to the jet. Such an enhanced dispersion can arise both from direct uplifting of the gas from the disc in the form of outflows \citep[e.g., see Fig.~14 of][]{meenakshi22}, as well as turbulence introduced into the gas disc by the jet driven bubble.

\section{Outflow scenarios}
\label{appendixC}

Here we describe the four scenarios that we considered for estimating the outflow masses and outflow mass rates, from the least to the most conservative.
For this analysis, we only used the CO(2-1) data at the original 0\farcs2 resolution to avoid artefacts or wrong interpretations of the results due to beam smearing of the larger CO(3-2) beam size. 

\textsc{i}) In the first scenario, we assume that all the molecular gas that cannot be reproduced with our rotating disc model is outflowing. Thus, we just integrated the emission the \barolo~model and subtracted it from the CO(2-1) datacube. We find that only 55\% of the CO(2-1) emission can be reproduced with rotation, and the remaining 45\% would correspond to an outflow mass of 5.1$\times$10$^8$\msol. However, as mentioned in Section \ref{sec:results}, the PVDs perpendicular to the jet (PA=-30\degree; bottom panels in Figure \ref{fig:pvd}) show an inverted S-shape in the central $r\sim$1\arcsec~(1.6\,kpc), which includes the bulk of the emission, and also low-velocity extended emission at r$>$1\arcsec~that is only detected at 2$\sigma_{\rm rms}$. Thus, there is some extended low-velocity gas that is not accounted for by our rotating model, and some nuclear high-velocity gas that is considered by \barolo~as rotation while it might not be. This uncertainty is not considered in the errors listed in Table \ref{tab:prop}. In this scenario, we adopted an outflow velocity of $v_{\rm out}$=100\kms, estimated from the mean velocity residuals from the \barolo~fit in {\color{blue} RA22}.

\begin{figure}[h]
 \resizebox{\hsize}{!}{\includegraphics{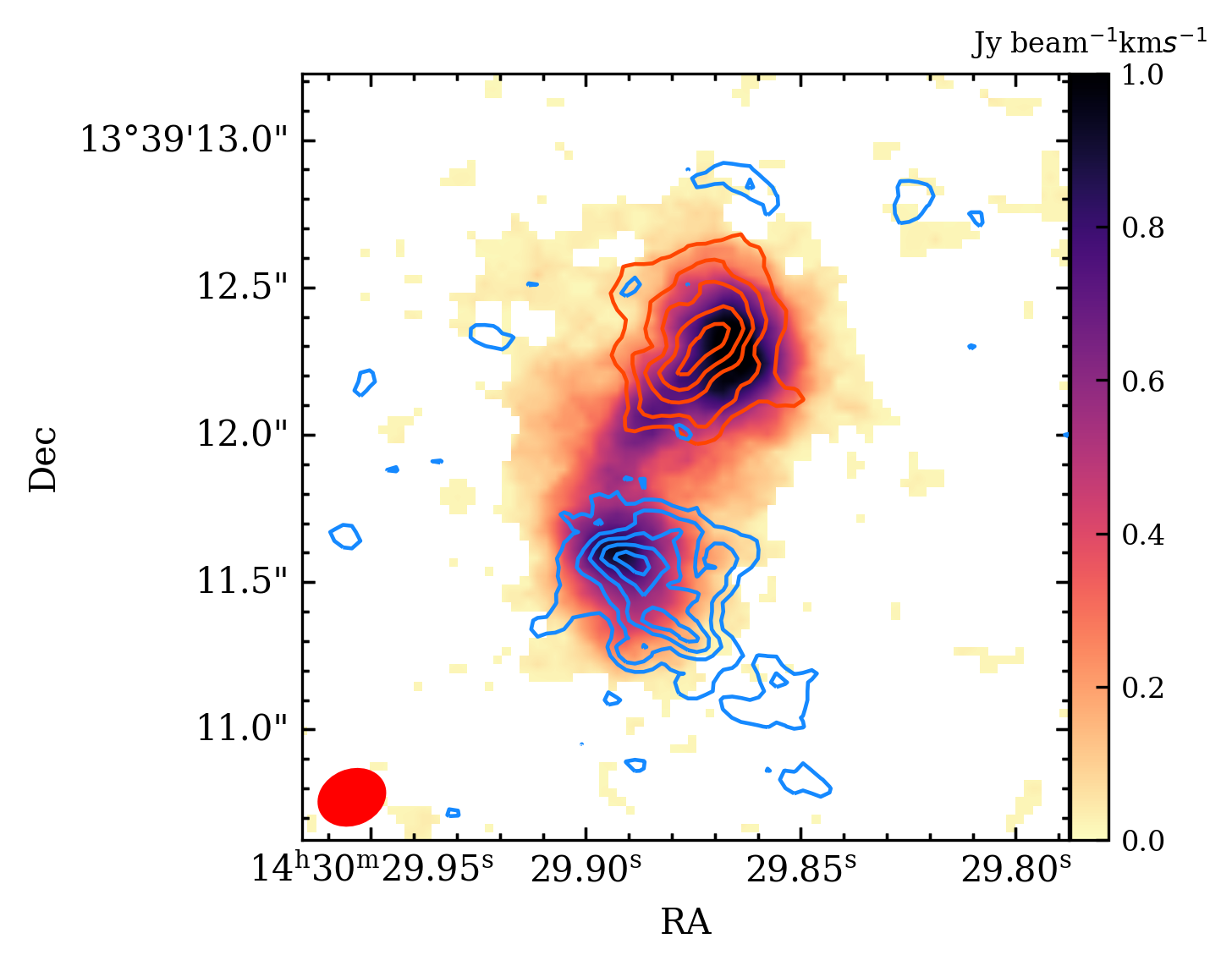}}
\caption{Integrated intensity of the high velocity components of the CO(2-1) emission at 0\farcs2 resolution. The colour map corresponds to the moment 0 of the total CO(2-1) emission, and the contributions from velocities faster than $v$=$\pm$300\kms~are shown as red and blue contours at (0.35,0.5,0.65,0.8,0.95)$\times\sigma_{\rm max}$, with $\sigma_{\rm max}^{\rm red}$= 0.28\,Jy\,beam$^{-1}$\,km\,s$^{-1}$ and $\sigma_{\rm max}^{\rm  blue}$=0.14\,Jy\,beam$^{-1}$\,km\,s$^{-1}$. The beam size is indicated with a red ellipse in the bottom left corner.}
\label{fig:high}
\end{figure}

\textsc{ii}) In the second scenario, we consider that only the highest velocities, faster than $\pm300$\kms, correspond to outflowing gas. We chose this value from the comparison of the PVDs shown in Figure \ref{fig:pvd} and the \barolo~rotation curve, shown in red the same figure. To calculate the flux of the high velocity gas, we created moment 0 maps by selecting only channels above $\pm$300\kms, shown in Figure~\ref{fig:high}. In this scenario, the mass corresponding to the outflow is 1.6$\times$10$^8$\msol.

\begin{figure}[h]
 \resizebox{\hsize}{!}{\includegraphics{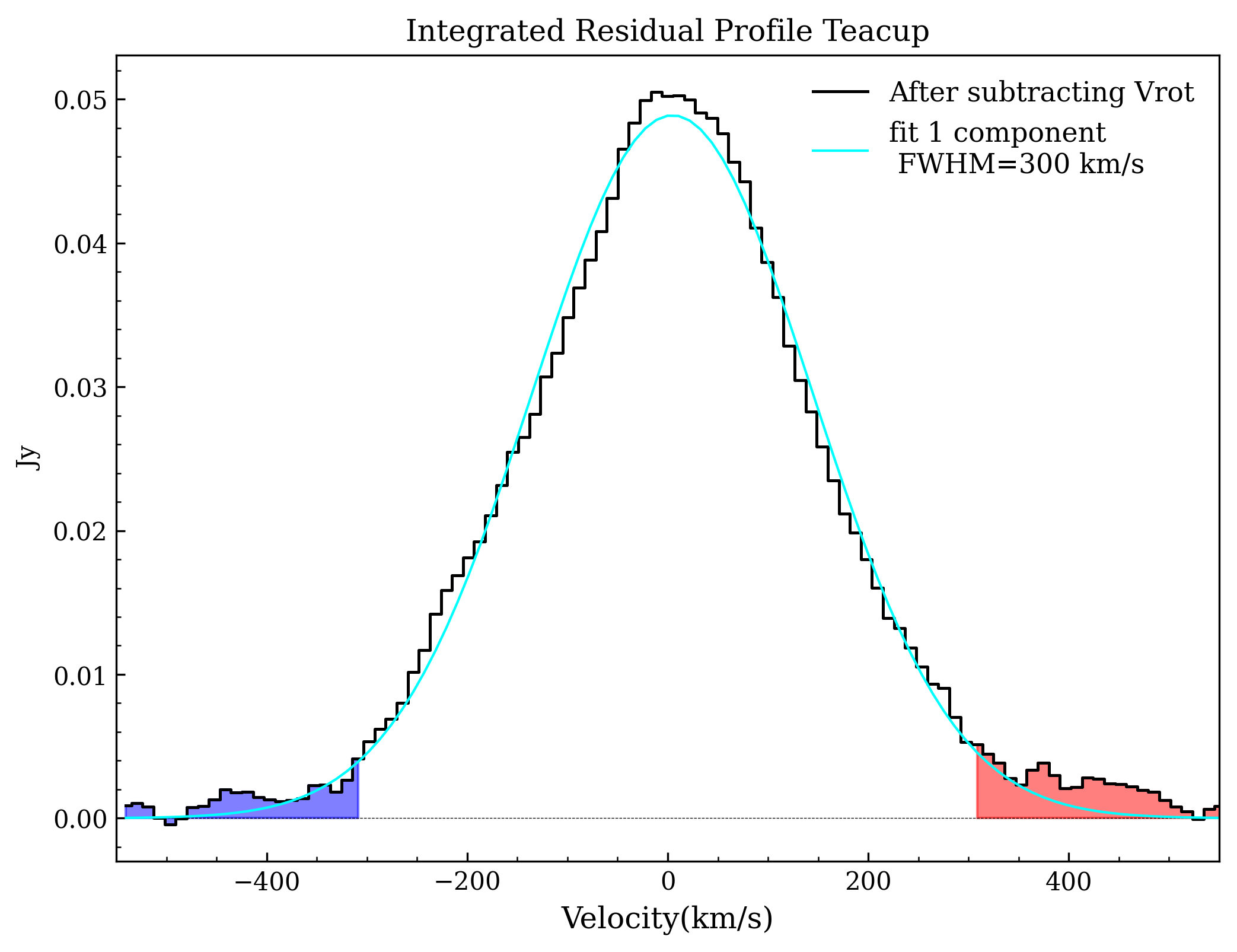}}
\caption{Integrated CO(2-1) residual profile after subtraction of the rotation curve. High-velocity blue and red wings are detected after subtraction. The cyan solid line corresponds to the fit with a single Gaussian component of FWHM$\sim$330\kms~used in Scenario \textsc{iv}. The blue and red shaded areas correspond to the high-velocity gas considered to compute the outflow flux in Scenario \textsc{iii}.}
\label{fig:res}
\end{figure}

\textsc{iii}) The third scenario is a mix of the first two. We assume that all the high-velocity gas is outflowing, but the contribution from rotation is subtracted following a similar approach as in \citet{santi19}. We deprojected the one dimensional rotation curve to the corresponding velocity field on a pixel by pixel basis, and re-shuffled the channels in order to remove the rotation component from the velocity axis of the CO(2-1) datacube. The result is a narrow residual CO profile, i.e., without the rotation component, which is shown in Figure~\ref{fig:res}. Since this method is suited to optimize the signal-to-noise of the emission associated to non-circular motions, it can reveal high-velocity wings that otherwise are too faint to be detected. Using the residual profile and considering only gas with velocities above $\pm$300\kms, we derive a mass of 5.7$\times$10$^7$\msol~for the outflow.

\textsc{iv}) The fourth scenario is based on the method described in scenario \textsc{iii}, but it assumes that the bulk of the residual CO profile shown in 
Figure~\ref{fig:res}, corresponds to turbulence unrelated to the outflow. This turbulence can be due to the merger, or produced by the jet, but here consider that it does not correspond to outflowing gas. The residual profile can be fitted with a single Gaussian of FWHM$\sim$330\kms, as shown in Figure~\ref{fig:res}. If we subtract this Gaussian from the total residual profile, we measure a mass of 2.6$\times$10$^7$\msol, which mainly corresponds to the high-velocity wings.

We computed the outflow masses from the integrated flux measurements in the outflow. First, the integrated fluxes are converted to CO luminosities L$_{\rm CO(2-1)}^\prime$, in units of ${\rm K\,km\,s^{-1}\,pc^2}$ using Equation 3 of \citet{solomon05} and then translated into masses using the CO-to-H$_2$ ($\alpha_{\rm CO}$) conversion factor M$_{\rm H_2}$=$\alpha_{\rm CO}$R$_{\rm 12}$L$^\prime_{\rm CO(2-1)}$. Under the assumption that the gas is thermalised and optically thick, the brightness temperature ratio $R_{\rm 12}={L_{\rm CO(1-0)}^\prime}/{L_{\rm CO(2-1)}^\prime}$=1. A conservative $\alpha$=0.8 is usually adopted for deriving outflow masses (see {\color{blue} RA22} and references therein). The values for each scenario are listed in Table~\ref{tab:prop}.

\begin{table}
\caption{Outflow measurements for the four scenarios proposed for the Teacup.}             
\label{tab:prop}      
\centering   
\tiny
\begin{tabular}{c c c c c c}         
\hline\hline  

Scenario & $S_{\rm CO}$ & $M_{\rm out}$ & $\dot{M}_{\rm out}$  & $M_{\rm out}/M_{\rm H_2}$  &$\dot{E}_{\rm kin}/P_{\rm jet}$\\  
          & (Jy\kms) & (10$^7$ M$_\odot$) & (\mout)  & (\%) &  \\
\hline                                  
\textsc{i}       & 7.6$\pm$0.9  & 51.5$\pm$48.2   &  44.0 $\pm$ 32.7$^\dag$   & 8.3 & 0.004\\
\textsc{ii}      & 2.4$\pm$0.1  & 16.0$\pm$10.9   &  41.0$\pm$27.9   & 2.6 & 0.035 \\
\textsc{iii}     & 0.9$\pm$0.1  & 5.6$\pm$4.3  &  15.1$\pm$11.0  &  0.9 & 0.013\\
\textsc{iv}      & 0.4$\pm$0.3  & 2.6$\pm$3.4  &  6.7$\pm$8.7 &  0.4 & 0.006 \\
\hline        
\end{tabular}
\\
\tablefoot{The uncertainty in $\alpha_{\rm CO}=$0.8$\pm$0.5\,\msol(K\,\kms\,pc$^2$)$^{-1}$ \citep{downes98} is included in the mass error estimates. $\dag$: computed using $v_{\rm out}$=100\kms, while for the others scenarios we adopted $v_{\rm out}$=300\kms. The outflow mass fractions were computed using the total mass of $M_{\rm H_2}$=6.2$\times 10^{9}$\msol~from {\color{blue} RA22}, estimated with $\alpha_{\rm CO}$=4.35\,\msol(K\,\kms\,pc$^2$)$^{-1}$.}
\end{table}

To compute outflow mass rates, we use an outflow radius of $r_{\rm out}$=0\farcs75 (1.2\,kpc), as constrained from the extent of the high-velocity gas with high $\sigma$ and \ratio~(see Figures \ref{fig:pvd} and \ref{fig:t32}). The outflow velocities reach up to $\pm$400\kms, but here we adopt a more conservative velocity of $v_{\rm out}$=300\kms, except for Scenario \textsc{i}. For a time-averaged thin expelled shell geometry \citep{rupke05}, $\dot{M}_{\rm out}$=$v_{\rm out} \left(M_{\rm out}/r_{\rm out}\right)$, which corresponds to the outflow mass averaged over the flow timescale, $t_{\rm flow}={\rm R_{\rm out}}/{\rm v_{\rm out}}$=3.9\,Myr. Considering all the previous, the outflow mass rates are in the range $\dot{M}_{\rm out}$=6.7$-$44 \mout. The ratio between the kinetic power of the molecular outflow ($\dot{E}_{\rm kin}=0.5 \dot{M}_{\rm out} {v_{\rm out}}^{2}$) and the jet power varies from 0.004$<\dot{E}_{\rm kin}/P_{\rm jet}<$0.035 (see Table~\ref{tab:prop}), indicating that the jet has enough kinetic energy to drive the molecular outflow.

\end{appendix}

\end{document}